\documentclass[letterpaper,aps,prl,showpacs,twocolumn, superscriptaddress,floatfix]{revtex4}
\usepackage{amsmath}
\usepackage{bm}
\usepackage{graphicx}
\usepackage{float}
\usepackage{array}
\usepackage{cancel}
\usepackage{color}
\usepackage{amsxtra}
\usepackage{amstext}
\usepackage{amssymb}
\usepackage{latexsym}
\usepackage{dsfont}
\usepackage{bbold}
\usepackage{isomath}

\usepackage{siunitx}
\usepackage{braket}
\usepackage{commath} 
\newcommand*\diff[1]{\mathop{}\!\mathrm{d^{#1}}} 
\DeclareMathOperator{\Tr}{Tr}
\newcommand{\ketbra}[2]{\ensuremath{\ket{#1}\!\bra{#2}}}

\newcommand{\E}[1]{\ensuremath{\operatorname{E}\left[#1\right]}}

\begin{document}

\title{Quantum Model of Cooling and Force Sensing With an Optically Trapped Nanoparticle}

\author{B.~Rodenburg*}
\affiliation{School of Physics and Astronomy, Rochester Institute of Technology, Rochester, NY 14623, USA}
\affiliation{Center for Coherence and Quantum Optics, University of Rochester, Rochester, NY 14627, USA}
\email{brandon.rodenburg@gmail.com}
\author{L.~P.~Neukirch}
\affiliation{Center for Coherence and Quantum Optics, University of Rochester, Rochester, NY 14627, USA}
\affiliation{Department of Physics and Astronomy, University of Rochester, Rochester, NY 14627, USA}
\author{A.~N.~Vamivakas}
\affiliation{Center for Coherence and Quantum Optics, University of Rochester, Rochester, NY 14627, USA}
\affiliation{Institute of Optics, University of Rochester, Rochester, NY 14627, USA}
\author{M.~Bhattacharya}
\affiliation{School of Physics and Astronomy, Rochester Institute of Technology, Rochester, NY 14623, USA}
\affiliation{Center for Coherence and Quantum Optics, University of Rochester, Rochester, NY 14627, USA}
\date{\today}
\begin{abstract}
    Optically trapped nanoparticles have recently emerged as exciting
    candidates for tests of quantum mechanics at the macroscale and as
    versatile platforms for ultrasensitive metrology. Recent experiments have
    demonstrated parametric feedback cooling, nonequilibrium physics, and
    temperature detection, all in the classical regime. Here we provide the
    first quantum model for trapped nanoparticle cooling and force sensing. In
    contrast to existing theories, our work indicates that the nanomechanical
    ground state may be prepared without using an optical resonator; that the
    cooling mechanism corresponds to nonlinear friction; and that the energy
    loss during cooling is nonexponential in time. Our results show excellent
    agreement with experimental data in the classical limit, and constitute an
    underlying theoretical framework for experiments aiming at ground state
    preparation. Our theory also addresses the optimization of, and the
    fundamental quantum limit to, force sensing, thus providing theoretical
    direction to ongoing searches for ultra-weak forces using levitated
    nanoparticles.
\end{abstract} 
\pacs{ 42.50.-p, 42.50.Wk, 37.10.Vz, 62.25.-g }

\maketitle

\section{Introduction}
\begin{figure*}[t!]
\centering
\includegraphics[width=\textwidth]{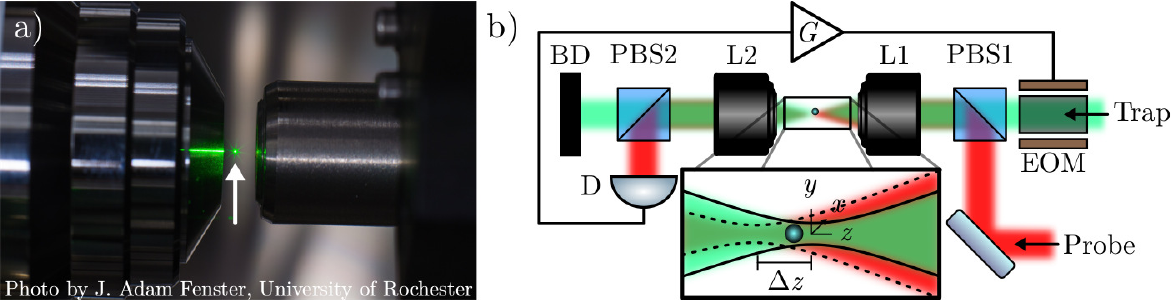}
\caption{a) Image of the trapped nanoparticle. b) Schematic of the experiment
    modeled in this article.}
\label{fig:P1}
\end{figure*}
Optically trapped nanoparticles can support explorations of macroscopic quantum
mechanics as well as ultrasensitive metrology very well since they can be
isolated from the environment in a trap, cooled, and detected - all using a
single laser beam without the need for an optical cavity~\cite{Neukirch2013,
Gieseler2014, Millen2014, Gieseler2015, Neukirch2015}. Experiments with
optically trapped harmonically oscillating subwavelength dielectric particles
~\cite{Li2011, Yin2013, Bateman2013, Arita2013, Scala2013} (see
Fig.~\ref{fig:P1}) have recently realized feedback
cooling~\cite{Gieseler2012,Neukirch2013}, nonlinear
dynamics~\cite{Gieseler2014}, non-equilibrium physics~\cite{Gieseler2015},
coupling to spin degrees of freedom~\cite{Neukirch2015a} and
thermometry~\cite{Millen2014}. All experiments thus far have been carried out
in the classical regime.

While several groups are currently exploring ways to access the nonclassical
regime of such systems, we present here the first quantum theory of trapped
nanoparticle optical feedback cooling and force sensing. The impetus for
investigating cooling comes from the fact that although levitated particles
have been successfully cooled in optical resonators (to
\SI{10}{K}~\cite{Millen2015} and \SI{64}{K}~\cite{Kiesel2013}), cavityless
cooling has been able to reach much lower temperatures $(50$ mK
\cite{Gieseler2012}) starting from the same initial (room) temperature. It is
however an open theoretical question as to whether the quantum ground state can
be prepared without using a cavity. Our model addresses this important question
and reveals a viable route to nanoparticle ground state preparation for ongoing
cavityless experiments. It also identifies the dissipative mechanism underlying
experimental cooling to be nonlinear in nature, in contrast to standard
experimental techniques, which depend on linear damping, and standard theory,
which relies on linear response analysis~\cite{Kippenberg2007, Marquardt2009,
Romero-Isart2011a, Aspelmeyer2014, Meystre2013, Kiesel2013, Asenbaum2013}.
Lastly, the model yields analytical results for the ensuing nonexponential
decay of phonon number, which shows excellent agreement with our experimental
data in the classical regime. The motivation for investigating force detection
is provided by the use of levitated cavityless nanoparticles in several ongoing
searches for various ultraweak forces~\cite{Geraci2010, Neukirch2015,
Ranjit2015, Moore2014}. Again it is an open question as to whether and to what
extent force detection is limited by the effects of quantum backaction in those
systems. To clarify this issue we derive in this article expressions for the
force sensitivity and the standard quantum limit of force detection. We expect
our new results on cooling and force sensing in the quantum limit will be
invaluable as this still nascent field matures.

A schematic of the physical system under consideration is shown in
Fig.~\ref{fig:P1}. A subwavelength polarizable dielectric sphere is confined at
the focus of a Gaussian trapping beam, and its motion is detected using a probe
beam, polarized orthogonal to the trap. The detected signal is processed and
fed back to the trap beam to cool the particle. We analyze this configuration
by dividing it into a `system' and a `bath'. The system consists of the
nanomechanical oscillator and the optical probe and trap. The bath consists of
the optical modes into which light is scattered by the nanosphere, and the
background thermal gas present in the experiment. We proceed to derive a
quantum model by identifying the electromagnetic modes relevant to the problem,
constructing the system and bath Hamiltonians, and deriving the master equation
for the system~\cite{Pflanzer2012, Carmichael2002}. All conclusions stated in
this article follow from this master equation.

\section{Model}
The configuration Hamiltonian can be written as,
\begin{equation}
\label{eqn:FullHamiltonian}
H = H_{m}+H_{f}+H_{\text{int}}.
\end{equation}
In Eq.~\eqref{eqn:FullHamiltonian}, the first term on the right hand side
represents the mechanical kinetic energy
$H_{m}=\left|\mathbf{p}\right|^{2}/2m$, where $\mathbf{p}$ is the three
dimensional momentum of the nanoparticle, and $m$ its mass. The second term in
Eq.~\eqref{eqn:FullHamiltonian} is the field energy $H_{f}=\epsilon_{0}\int
\left|\mathbf{E(r)}\right|^{2}\diff3\mathbf{r}$, where $\mathbf{E(r)}$ is the
sum of the trap $\mathbf{E_t}$, probe $\mathbf{E_p}$, and background
$\mathbf{E_b}$ electric fields. We model the trap and probe modes as Gaussian
beams, and the background using a plane wave expansion. We find after dropping
a constant term,
\begin{equation}
\label{eqn:FreeFieldHamiltonian2}
H_{f}=\hbar \omega_{p}a^{\dagger}a+\sideset{}{}\sum_{\mu}\int \diff3\mathbf{k}\hbar
\omega_{\mathbf{k}}a^{\dagger}_{\mu}(\mathbf{k})a_{\mu}(\mathbf{k}),
\end{equation}
which is simply the sum of the probe and background field energies with $a$ and
$a_\mu(\mathbf k)$ representing the corresponding standard bosonic field
operators. Finally, in Eq.~\eqref{eqn:FullHamiltonian} the interaction
Hamiltonian is given by $H_{\text{int}} = -\int_{V} \mathbf P(\mathbf r)\cdot
\mathbf E(\mathbf r)\diff3\mathbf{r}/2= -\alpha_p\int_{V} \left|\mathbf
E(\mathbf r)\right|^{2}\diff3\mathbf{r}/2$, where we have assumed that the
dielectric has a volume $V$, and that it has a linear polarizability density
$\alpha_p$, i.e. the polarization density is $\mathbf P(\mathbf r) = \alpha_p
\mathbf E(\mathbf r)$. Using the expressions for the electric fields from the
Supplementary Material we can evaluate $H_{\text{int}}$ for small particle
displacements $\mathbf{r}$, and rewrite Eq.~\eqref{eqn:FullHamiltonian} as
\begin{equation}
\label{eqn:HamRedo}
H=H_{S}+H_{B}+H_{SB},
\end{equation}
where this system Hamiltonian is 
\begin{equation}
\label{eqn:SystemHamiltonian}
H_{S}=\hbar \omega_{p}a^{\dagger}a +\sum_{j}\hbar \omega_{j}b^{\dagger}_{j}b^{}_{j}
-\sum_{j}\hbar g_{j} a^{\dagger}a(b^{}_{j}+b^{\dagger}_{j}),
\end{equation}
with mechanical trapping frequencies $\omega_j$, optomechanical coupling
constants $g_j$, and mechanical operators which obey the standard commutation
relations $[b^{}_{j},b^{\dagger}_{j}]=1$ ($j = \left\{x,y,z\right\}$).
In Eq.~\eqref{eqn:HamRedo}, the bath
Hamiltonian is $H_{B}=\sum_{\mu}\int \diff3\mathbf{k}\hbar
\omega_{\mathbf{k}}a^{\dagger}_{\mu}(\mathbf{k}) a_{\mu}(\mathbf{k})$ and the
system-bath interaction Hamiltonian is
$H_{SB}=-\epsilon_{c}\epsilon_{0}\int_{V}\diff3\mathbf{r}[\mathbf{E_{t}(r)}+\mathbf{E_{p}(r)}]
\cdot \mathbf{E_{b}(r)}$, which represents the scattering of the trap and probe
fields into the background (see Supplementary Material for details).

We now trace over the bath modes, applying the standard Born and Markov
approximations, since the system-bath coupling is weak and the bath
correlations decay quickly~\cite{Romero-Isart2011a, Pflanzer2012}. We also
trace over the $x$ and $y$ degrees of particle motion,
since the dynamics along the three axes are independent of each other, and it
suffices to analyze a single direction~\cite{Gieseler2012}. The net result of
our calculation is a master equation for the density matrix $\rho(t)$
describing the optical probe and the $z$-motion of the nanoparticle
\begin{equation}
\label{eqn:Master1}
\dot{\rho}(t) = \frac{1}{i\hbar}[H_{S}', \rho]
                - \frac{A_t}{2}\mathcal D[Q]\rho
                + \mathcal{L}_{\text{sc}}\rho,
\end{equation}
where the first term on the right hand side represents unitary evolution of the
system with $H_{S}' = \hbar \omega_{p}a^{\dagger}a + \hbar
\omega_{z}b^{\dagger}_{z}b_{z} - \hbar g_{z}
a^{\dagger}a(b_{z}+b^{\dagger}_{z})$. The second term corresponds to the
positional decoherence of the nanoparticle due to scattering of trap photons,
with the Lindblad superoperator $\mathcal D[Q_z]\rho \equiv
Q_z^\dagger Q_z\rho+ \rho Q_z^\dagger Q_z - 2Q_z\rho Q_z^\dagger$, where $Q_z =
b^{\dagger}_{z}+b_{z},$ and $A_t$ is the heating rate due to trap beam
scattering as defined in the Supplementary Material. The third superoperator
describes the loss of photons from the probe, also due to scattering by the
nanoparticle, $\mathcal{L}_{\text{sc}}[\rho(t)] = -B\left(\mathcal D[a] +
(7\omega_p^2 \ell_{z}^{2}/5c^2)\mathcal D[aQ_z]\right)\rho$, where
$\ell_{z}$ is the oscillator length.

The nanoparticle also experiences collisions with background gas particles at
the ambient temperature $T$. This effect may be accounted for by adding to the
right hand of Eq.~\eqref{eqn:Master1} the superoperator~\cite{Diosi1995}
\begin{equation}
\begin{split}
\mathcal{B}[\rho(t)]
    = & -\frac{D_p}{2}\mathcal D[Q_z]\rho - \frac{D_q}{2}\mathcal D[P_z]\rho\\
      & -i\frac{\eta_f}{4m}\left[Q_z,\left\{P_z,\rho\right\}\right],
\end{split}
\label{eqn:BrownianTerms}
\end{equation}
where $P_z = i(b^\dagger_z - b_z)$, and curly braces denote an anticommutator.
The first term on the right hand side corresponds to momentum diffusion and
$D_p=2\eta_fk_BT\ell_z^2/\hbar^2$, where $k_B$ is Boltzmann's constant. The
second term describes position diffusion with $D_q =
\eta_f\hbar^2/(24k_BTm^2\ell_z^2)$. The third term accounts for friction, and
by Stokes law we have $\eta_f = 6\pi\mu r_d$, where $r_d$ is the radius of the
nanoparticle and $\mu$ is the dynamic viscosity of the background gas. As shown
earlier, internal and center-of-mass heating of the nanoparticle due to optical
absorption and blackbody radiation are negligible in systems such as ours, as
are particle size and shape effects, as well as trap beam shot
noise~\cite{Gieseler2012, Chang2010}.

We now characterize the measurement of the oscillator displacement using input-output
theory from quantum optics ~\cite{Gardiner2004} applied to the nanoparticle.
Specifically, the incoming probe field $a_{\text{in}}$ interacts with the nanoparticle, and the
outgoing probe field $a_{\text{out}}$ carries a signature of this
interaction (as shown in the Supplementary Material)
\begin{equation}
a_{\text{out}}=a_{\text{in}}+\frac{\alpha \chi}{2} Q_{z}(t),
\end{equation}
where $\chi = 4 g_z\Delta t$ is the scaled optomechanical coupling, with
integration time $\Delta t$ (determined by the detection bandwidth),
and we have written the probe beam as a coherent state $a =-i\alpha+v$, with
$\alpha$ a classical number and $v$ a bosonic annihilation operator. A
homodyne measurement on the output field yields a current~\cite{Gardiner2004}
\begin{equation}
\label{eqn:HomodyneCurrent}
I_{h}=\chi^2\Phi\langle Q_{z}\rangle (t)+\sqrt{\chi^2\Phi}\xi(t),
\end{equation}
where $\Phi=\alpha^{2}\Delta\omega$ is the average detected flux of probe
photons, and $\xi (t)$ is a stochastic variable with mean $\langle \xi(t)\rangle=0$
and correlation $\langle \xi(t)\xi(t')\rangle=\delta(t-t')$.

\begin{figure}[th]
\centering
\includegraphics[width=0.45\textwidth]{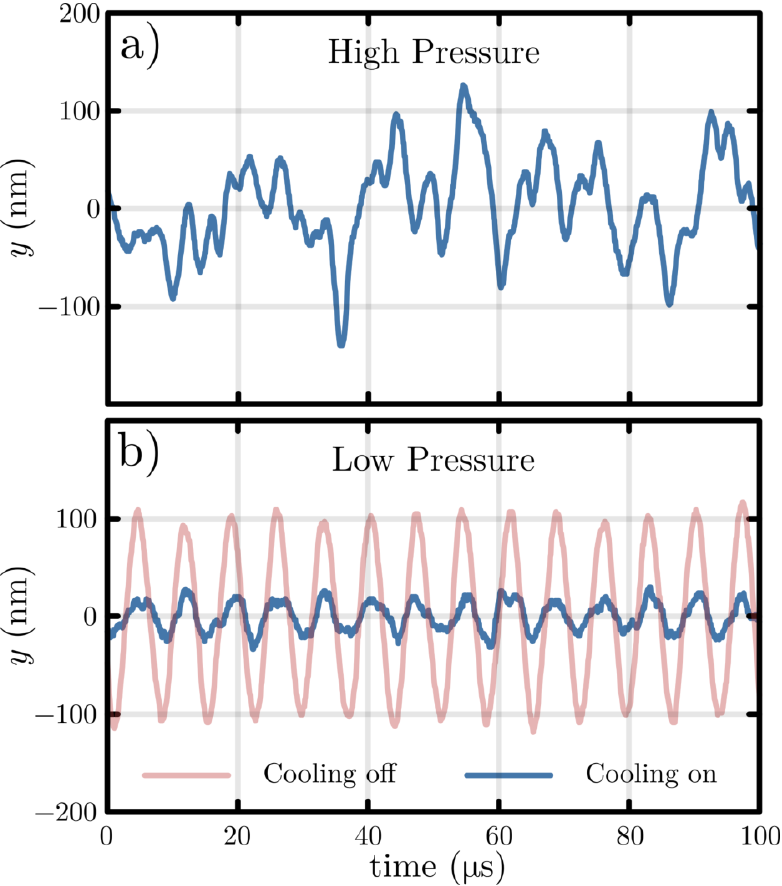}
\caption{a) Shows the diffusive evolution of the trapped
    nanoparticle's position at atmospheric pressure. b) Harmonic motion
    of the nanoparticle at a lower pressure of \SI{4e-3}{mbar}. The reduction of the
    amplitude of harmonic motion corresponds to the turn-on of feedback, i.e. cooling.}
\label{fig:y(t)}
\end{figure}

In the experiment, the detected current $I_h$ is frequency doubled, phase
shifted, and fed back to modulate the power of the trapping
beam~\cite{Gieseler2012}. This results in a feedback Hamiltonian
$H_\text{fb}=\hbar G I_\text{fb}Q^3_z$, where $G$ is the dimensionless feedback
gain related to the trap intensity modulation~\cite{Neukirch2015,
Mancini1998},
\begin{equation}
    M\equiv \frac{\Delta I_t}{I_t} \approx \frac{G\chi^2\Phi\braket{b_z^\dagger b_z}}{\omega_z},
\label{eqn:TrapMod}
\end{equation}
and the feedback current is $I_\text{fb} =
\chi^2\Phi\braket{P_z}+\sqrt{\chi^2\Phi}\xi'(t)$, where $\xi'(t)$ has the same
properties as $\xi(t)$. This form of the Hamiltonian implies a feedback force
$F_{\text{fb}}=-\partial H_{\text{fb}}/\partial Q_{z}$ which is equivalent to
that used in experiments in the classical regime~\cite{Neukirch2015}. Taking
the Markovian limit where the feedback occurs faster than any system timescale,
and applying quantum feedback theory for homodyne
detection~\cite{Wiseman1993a}, we find that the following superoperator must be
added to Eq.~\eqref{eqn:Master1}
\begin{equation}
\label{eqn:Feedback1}
\mathcal{F}[\rho(t)]=-i\chi^2\Phi G[Q_{z}^{3},\{P_{z},\rho\}]
-\frac{\chi^2\Phi}{2}G^{2}\mathcal D[Q_z^3]\rho,
\end{equation}
where the first term on the right hand side represents the desired cooling
effect of the feedback, and the second term the accompanying backaction. We
emphasize that in contrast to standard optomechanics, the feedback and
backaction terms are highly nonlinear in the oscillator variables. The presence
of this nonlinearity distinguishes our system from conventional cavity
optomechanics and results in qualitatively different dynamics, as we show
below.

\begin{figure*}[t!]
\centering
\scriptsize
\tiny
\includegraphics[width=\textwidth]{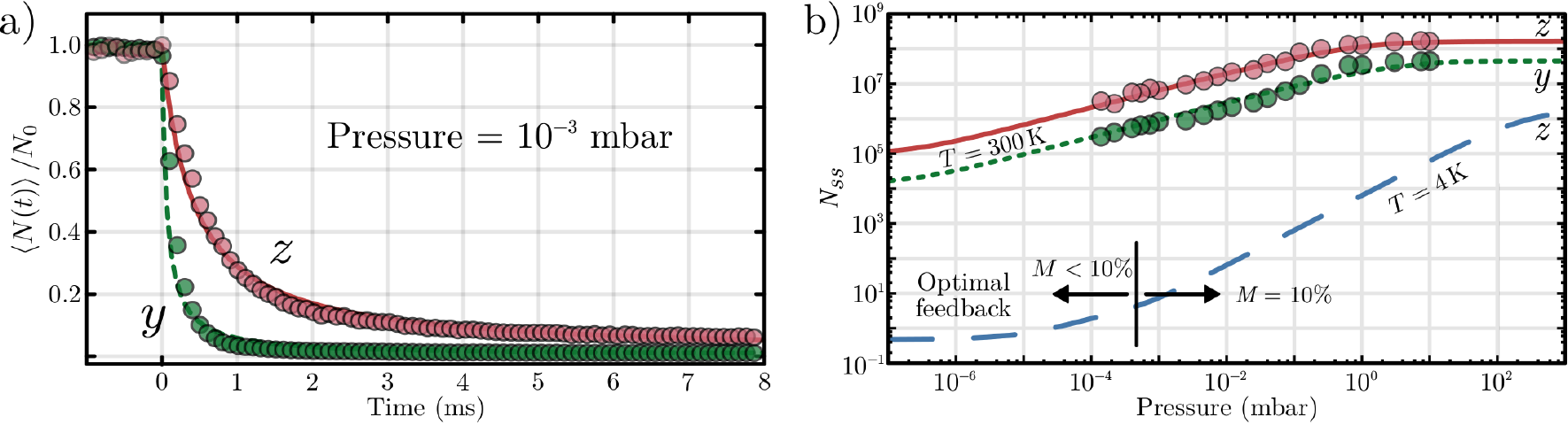}
\caption{a) The $y$ and $z$ phonon cooling dynamics
    [Eq.~\eqref{eqn:DynamicsSolution}]. b) Steady state phonon number versus
    pressure [Eq.~\eqref{eqn:Nss_exact}]. Circles represent experimental data,
    and the solid curve our theoretical model for a fused silica sphere
    ($\epsilon_r = 2.1$ and $\text{density} = \SI{2200}{kg\per m^3}$) radius
    $r_d=\SI{50}{nm}$, \SI{1064}{nm} trap (\SI{100}{mW}) and probe
    (\SI{10}{mW}) beams, a mechanical frequency $\omega_{z}/2\pi =
    \SI{38}{kHz}$, $\chi \approx \num{e-7}$, and trap intensity modulation $M
    \lesssim 0.1\%$. The dotted lines represent the equivalent curves for one
    of the transverse degrees of freedom ($\omega_y/2\pi = \SI{138}{kHz}$). The
    dashed curve in a) represents the prediction of our theory for a setup
    placed in a cryostat with the feedback chosen optimally, keeping $M\leq
    10\%$.}
\label{fig:DataPlots}
\end{figure*}

The full master equation, assembled from Eqs.~\eqref{eqn:Master1},
\eqref{eqn:BrownianTerms} and~\eqref{eqn:Feedback1} is then
\begin{equation}
\begin{split}
\dot{\rho}(t)
    &= \frac{1}{i\hbar}[\tilde{H}_{S}, \rho(t)]
    - (A_t + A_p)\mathcal D[Q_z]\rho(t)/2
\\&\qquad
    - B\mathcal D[a]\rho(t)
    + \mathcal{B}[\rho(t)]
    + \mathcal{F}[\rho(t)],
\end{split}
\label{eqn:ME}
\end{equation}
where the new system Hamiltonian $\tilde{H}_{S} = \hbar \omega_pv^\dagger v +
\hbar \omega_zb^\dagger_zb_z - i\alpha\hbar g_z(v^\dagger - v)(b_z +
b^\dagger_z)$, accounts for the linearization of the probe implemented above,
and the master equation now includes a mechanical decoherence term due to
scattering from the probe, in addition to the trap beam, with heating rate
$A_p$.

A sample experimental data set of the measured position of the nanoparticle
along the $y$ axis is shown in Fig.~\ref{fig:y(t)}. At atmospheric pressures,
Brownian effects, as given by Eq.~\eqref{eqn:BrownianTerms} dominate, and the
position of the particle follows a diffusive evolution, as can be seen in
Fig.~\ref{fig:y(t)}a. At lower pressures the particle's evolution becomes
increasingly ballistic. The ensuing harmonic motion is shown in
Fig.~\ref{fig:y(t)}b, both in the absence as well as in presence of feedback
[Eq.~\eqref{eqn:Feedback1}]. The decrease in amplitude of the harmonic motion
is due to the presence of parametric feedback cooling.

\section{Phonon dynamics}
Employing the master equation [Eq.~\eqref{eqn:ME}] to consider the question of
ground state occupation, tracing out the optical probe field, and using the
resulting reduced master equation for the nanoparticle only, we find the
equation for the dynamics of the phonon number $(N \equiv
b^{\dagger}_{z}b_{z})$, $\langle \dot{N}\rangle = -J\langle N^{2}\rangle -
K\langle N\rangle + L,$ where $J = [12G - 54G^2]\chi^2\Phi$, $K = \eta_{f}/2m +
J$, $L=D - J/2$, the dot denotes a time derivative, and $D=D_{p}'+D_{q}$ with
$D_{p}'=D_{p}+A_{t}+A_{p}$ accounting for positional decoherence. We assume
that the nanoparticle is described by a thermal state~\cite{Wilson-Rae2007,
Marquardt2007, Genes2008, Romero-Isart2011b, Pflanzer2012}, for which $\langle
N^{2}\rangle = 2\langle N\rangle^{2}+\langle N\rangle$~\cite{Gerry2004}, a
relation which simplifies the phonon dynamics to
\begin{equation}
\label{eqn:PhononDynamics}
\langle \dot{N}\rangle = -2J\langle N\rangle^{2}-(J+K)\langle N\rangle+L.
\end{equation}
The total effect of parametric feedback on the phonon dynamics is contained in
the parameter $J$, which is determined by the difference between the feedback
cooling and backaction heating. In the experiments $J \neq
0$~\cite{Gieseler2012,Neukirch2013}, making the phonon dynamics of cooling
nonlinear, and the oscillator energy loss nonexponential, as shown below. We
stress that this behavior is \textit{qualitatively} different from standard
quantum cavity optomechanical theory, which characterizes cooling as a linear
damping process resulting in an exponential decay of energy (see,
e.g.~\cite{Wilson-Rae2007, Marquardt2007} and Eq. (82)
in~\cite{Aspelmeyer2014}). We note that $G = G_\text{opt} = 1/9$ maximizes
$\braket{\dot N}$ in Eq.~\eqref{eqn:PhononDynamics} with the maximum nonlinear
cooling rate $J_\text{max} = 2\chi^2\Phi/3$.

Assuming the initial condition $\langle N(0)\rangle \equiv
N_{0}=D_p'2m/\eta_f\equiv k_{B}T_\text{eff}/\hbar\omega_{z},$ where
$T_\text{eff}$ is the effective temperature of a bath due to gas and optical
scattering combined, the analytical solution  to Eq.~\eqref{eqn:PhononDynamics}
is
\begin{equation}
    \braket{N(t)} = -\frac{(J+K)}{4J}
        + \frac{1}{2J\tau}\tanh\left(\frac{t}{\tau}+ \theta\right),
\label{eqn:DynamicsSolution}
\end{equation}
where $\theta=\tanh^{-1}\left[(2JN_0+J+K)\tau\right]$ and the cooling timescale
$\tau = 2\left[(J+K)^2 + 8JL\right]^{-1/2}$. From
Eq.~\eqref{eqn:DynamicsSolution} the steady state phonon number is
\begin{subequations}
\begin{align}
N_\text{ss} \equiv \lim_{t\to\infty}\braket{N(t)}
    = \frac{1}{2J\tau}-\frac{(J+K)}{4J}
\label{eqn:Nss_exact}
    \\ \approx \sqrt{\frac{\eta_f}{2m}\frac{N_0}{2J}}
    = \sqrt{\frac{D_p+A_t+A_p}{2J}},
\end{align}
\label{eqn:SteadyStatePhononNumber}%
\end{subequations}
where the approximation is valid for $N_0\gg 1$. To reach the ground state, we
need to maximize the feedback cooling $J$, which can be done by setting
$G=G_\text{opt}$. We also need to minimize gas heating, which can be
accomplished by going to low pressures and cryogenic temperatures, such that
$D_p$ is negligible in Eq.~\eqref{eqn:SteadyStatePhononNumber}. Below we discuss
situations involving realistic experimental parameters.

Two plots of the nonlinear phonon dynamics are shown in
Fig.~\ref{fig:DataPlots}a for the $z$ and $y$ motion at \SI{e-3}{mbar} along
with experimental values measured by us (circles). The solid curve represents
$\braket{N(t)}$ as given in Eq.~\eqref{eqn:DynamicsSolution}, while the dotted
curve gives the corresponding equation for motion along one of the transverse
directions $y$ (which is nearly degenerate with $x$ i.e.~$\omega_y \approx
\omega_x$), see Fig.~\ref{fig:P1}b. In Fig.~\ref{fig:DataPlots}b we show three
plots of the steady state phonon number, as the vacuum pressure is tuned. The
solid and dotted curves represent $N_\text{ss}$ [Eq.~\eqref{eqn:Nss_exact}] at
\SI{300}{K} for $z$ and $y$ motion respectively, while the circles are
experimental data. As can be seen, in all cases there is very good agreement
between theory and experiment. The dashed curve in Fig.~\ref{fig:DataPlots}b
predicts the steady state phonon number for an identical configuration, but
placed in a cryostat at \SI{4}{K}. The ground state can be prepared if starting
at high pressures, the particle is cooled while continuously increasing the
feedback gain as in~\cite{Gieseler2012}, and keeping the trap modulation $M =
10\%$. Proceeding in this manner, we find that below $\lesssim\SI{e-5}{mbar}$
optimal feedback $J = J_\text{max}$ can be achieved, and the ground state
occupied.

We note that practical cooling to lower phonon numbers is currently limited by
a number of factors. These include high pressures enforced by nanosphere
loading technologies, classical errors from the electronic feedback loop and
laser noise, measurement uncertainties due to detector bandwidth limitations,
and collection inefficiencies of the scattered light~\cite{Gieseler2012,
Millen2014, Neukirch2015, Ranjit2015}. However, these problems are technical
rather than fundamental, and efforts are underway to overcome these
limitations~\cite{Mestres2015}, suggesting strongly the possibility of cooling
to the ground state using the parameters presented in this article.

\begin{figure}[th]
\centering
\includegraphics[width=0.45\textwidth]{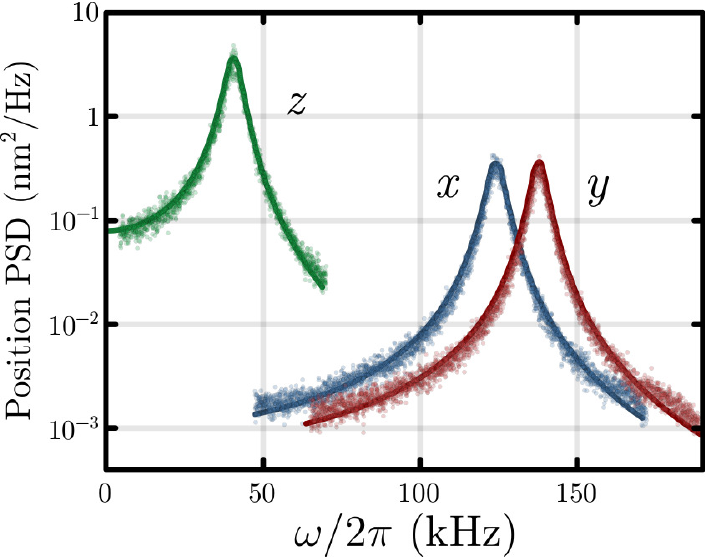}
\caption{Experimentally measured positional PSDs for all three degrees of
    freedom with dark lines representing the theoretical fits to the data
    [Eq.~\eqref{eqn:qPSD}]. Data was taken at a moderate vacuum pressure of
    \SI{10}{mbar} and clearly shows the Lorentzian shape of the
    resonance. These fits were used to extract the values of $\omega_j$,
    $\Gamma$, and $N_\text{ss}$.}
\label{fig:PSDs}
\end{figure}

\section{Force sensing}
We now consider force sensing using the nanoparticle model given by the master
equation of Eq.~\eqref{eqn:ME}. Since the state of the nanoparticle is
continuously monitored, the master equation can be unraveled in terms of a set
of Langevin equations describing the evolution of the quadratures $Q_z$ and
$P_z$ plus a stochastic force due to the measurement backaction which
gives~\cite{Gisin1992, Halliwell1995}
\begin{equation}
\begin{split}
\dot Q_z &= \mathcal L_0^\#[Q_z] = \omega_z P_z\\
\dot P_z &= \mathcal L_0^\#[P_z] + F/m\omega_z\ell_z\\
         &= -\omega_z Q_z - \Gamma P_z + F/m\omega_z\ell_z,
\end{split}
\label{eqn:Langevin}
\end{equation}
where $\mathcal L_0^\#$ is the Liouvillian superoperator (dual to the
superoperator $\mathcal L_0$ appearing in the master equation $\dot \rho =
\mathcal L_0[\rho]$) defined by $\Tr(\rho\mathcal L_0^\#[A]) \equiv
\Tr(A\mathcal L_0[\rho])$ for any arbitrary operator $A$~\cite{Hornberger2009}.
The parameter $\Gamma = \Gamma_0 + \delta\Gamma$, where $\Gamma_0$ is the gas
damping and $\delta\Gamma \approx 12\chi^2\Phi G(\braket{N} + 1/2)$ is the
nonlinear feedback damping~\cite{Gieseler2012, Neukirch2015, Gieseler2014}.
Finally $F = F_T + F_F$ is the sum of the (independent) stochastic forces due
to thermal and feedback backaction heating, respectively, with zero mean and
correlations $\Braket{F_T(t)F_T(t')} = S_T\delta(t-t')$ and
$\Braket{F_F(t)F_F(t')} = S_F\delta(t-t')$, with
\begin{equation}
\begin{split}
S_T &= 2m\Gamma_0k_BT_\text{eff}\\
S_F &= 27m\hbar\omega_z\chi^2\Phi G^2 \left(2\braket{N}^2 + 2\braket{N} + 1\right).
\end{split}
\label{eqn:STSF}
\end{equation}
The presence of the $\braket{N}$-dependent factor in Eq.~\eqref{eqn:STSF}
implies that the feedback noise is dependent on the system state, and is
therefore non-additive. Furthermore, the dependence is nonlinear in
$\braket{N}$. Both of these features are fundamentally different from the
typical additive feedback noise in standard cavity optomechanics, which is
independent of the state of the system~\cite{Genes2008}.

We convert Eq.~\eqref{eqn:Langevin} into the second order differential
equation for the position $\ddot q_z + \Gamma\dot q_z + \omega_z^2 q_z = F/m$, and take its
Fourier transform to find the position spectrum $\tilde q_{z}(\omega) =
\chi_m(\omega)\tilde F(\omega)$, where
\begin{equation}
\chi_m(\omega) = \left\{m\left[\left(\omega_z^2-\omega^2\right) - i\omega\Gamma\right]\right\}^{-1},
\end{equation}
is the optomechanical susceptibility of our oscillator. Finally the
positional power spectral density (PSD) noise spectrum is given by
\begin{equation}
\Braket{\abs{\tilde q_{z}(\omega)}^2}
= \abs{\chi_m}^2\left(S_T + S_F \right) + \frac{\ell_z^2}{\chi^2 \Phi},
\label{eqn:qPSD}
\end{equation}
where the last term in the equation comes from the shot noise of the measured
signal [Eq.~\eqref{eqn:HomodyneCurrent}]. A typical example data set of the
positional PSDs at moderate vacuum is shown in Fig.~\ref{fig:PSDs}, along with
fits to the theoretical expression of Eq.~(\ref{eqn:qPSD}). 

\begin{figure}[th]
\centering
\includegraphics[width=0.45\textwidth]{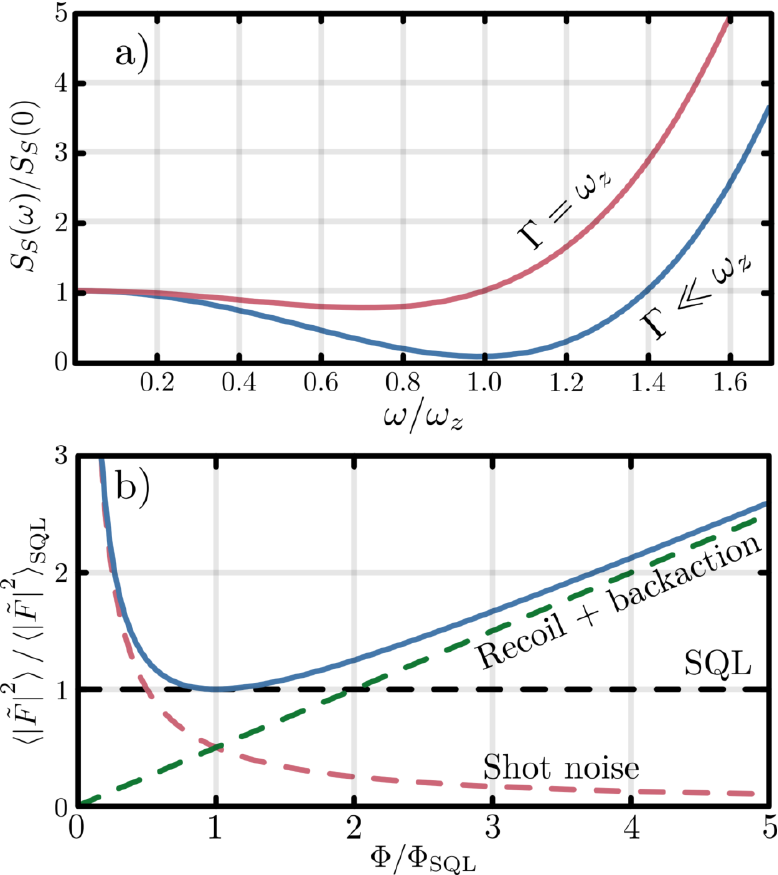}
\caption{a) Plots of the shot noise force PSD [the last term of
    Eq.~\eqref{eqn:FPSD}] versus the normalized mechanical frequency
    $\omega/\omega_{z}$ for low and high total damping $\Gamma$. The minimum
    occurs for $\omega_\text{opt} = \sqrt{\omega_z^2 - \Gamma^2/2}$. b) Plot of
    the force sensitivity as a function of the normalized optical power
    $\Phi/\Phi_\text{SQL}$ at high vacuum. The standard quantum limit is
    reached when the shot noise balances the recoil and backaction noises
    [Eq.~\eqref{eqn:FPSD}].}
\label{fig:ForceSensitivity}
\end{figure}

In view of the fact that trapped nanoparticles offer the possibility of
ultrasensitive force measurements~\cite{Moore2014, Neukirch2015, Ranjit2015,
Geraci2010}, we express our measurement noise spectrum
[Eq.~\eqref{eqn:qPSD}] in terms of the estimator $\tilde q(\omega)/\chi_m$
in order to investigate the fundamental limits of such measurements.
The sensitivity of force estimation is set by the force noise PSD
\begin{equation}
\Braket{\abs[1]{\tilde F(\omega)}^2}
    = S_T + S_F + S_S(\omega),
\label{eqn:FPSD}
\end{equation}
where $S_S(\omega) = S_S(0) \left[ \left(1-(\omega/\omega_z)^2\right)^2
+(\omega\Gamma/\omega_z^2)^2 \right]$ and $S_S(0) =
(m\ell_z\omega_z^2)^2/\chi^2\Phi$. Only the last term carries an $\omega$
dependence in Eq.~\eqref{eqn:FPSD}. A plot of $S_S(\omega)$ is shown in
Fig~\ref{fig:ForceSensitivity} in the high as well as low total damping
$\Gamma$ regimes, both of which are experimentally accessible
\cite{Gieseler2012,Neukirch2015a}. The minimum value of $S_S(\omega)$, and
therefore the optimal force sensitivity, occurs at the response frequency
$\omega_\text{opt} = \sqrt{\omega_z^2 - \Gamma^2/2}$.

The first two terms in Eq.~\eqref{eqn:FPSD} scale linearly with the optical
power while the shot noise scales inversely (i.e.~$S_T + S_F \sim \Phi$ and
$S_S \sim 1/\Phi$). Therefore there is a power that minimizes the total noise,
representing the standard quantum limit for our system. Assuming that the
feedback is optimal (i.e.~$J = J_\text{max}$), the standard quantum limit is
reached when $\chi^2\Phi_\text{SQL} \approx \Gamma/8$ and equals
\begin{equation}
\Braket{\abs[0]{\tilde F}^2}_\text{SQL}
    = 2m\Gamma_0k_BT + 4m\hbar\omega_z\left(A_t + \sqrt{2}\Gamma\right).
\label{eqn:FPSQL}
\end{equation}
The first term in \eqref{eqn:FPSQL} represents a thermal contribution from the
background gas; the second term is due to scattering of photons from the
trapping beam; and the third term contains the effects of light scattering and
shot noise from the probe, as well as the feedback backaction. At the low
vacuum pressures currently available (i.e.~$\lesssim\SI{e-7}{mbar}$) the gas
contribution is negligible, implying a minimum force sensitivity of
$\sqrt{\braket{\abs[0]{\tilde F}^2}_\text{SQL}} \approx \SI{e-21}{N\per\sqrt{Hz}} = \si{zN\per\sqrt{Hz}}$ and optimal probe power of
$\hbar\omega_p\Phi_{SQL} \approx \SI{2}{mW}$, where the remaining system
parameters have been taken from the caption of Fig.~\ref{fig:DataPlots}. Even
at this limit the system can be readily used to test for violations of
Newtonian gravity ($\sim \SI{e-18}{N}$)~\cite{Ranjit2015} with moderate
measurement bandwidths.  However, backaction effects will impose long
interrogation times on experiments searching for new small scale ($\sim
\SI{e-21}{N}$)~\cite{Geraci2010}, and Casimir forces ($\sim
\SI{e-24}{N}$)~\cite{Geraci2010, Neukirch2015}.  Conversely for short
measurement times, our calculations show that backaction effects, which are of
interest in their own right in optomechanics~\cite{Purdy2013}, can be observed
at moderate laser powers and readily attainable vacuum pressures.

\section{Conclusions}
To conclude, we have presented a quantum model that describes the cooling and
force sensing characteristics of an optically trapped subwavelength dielectric
particle. We have shown that the predictions of this model for cooling are in
very good agreement with experimentally measured occupation values in the
classical regime. Further, we have demonstrated that quantum ground state
preparation is challenging, but achievable in anticipated experiments. Finally,
we have derived the standard quantum limit to force sensing, indicating
experiments where the role of quantum backaction needs to be accounted for.
The model presented by us opens the door to the characterization of the quantum
behavior of a system important for macroscopic quantum
mechanics~\cite{Gieseler2012, Neukirch2013}, optical
tweezing~\cite{Gieseler2014}, ultrasensitive metrology~\cite{Millen2014}, and
non-equilibrium physics~\cite{Gieseler2015}. With the proper identifications,
our theory is also applicable to electromechanical systems with parametric
feedback~\cite{Villanueva2011}.

\section{Acknowledgements}
We are grateful to C.~Stroud, A.~Aiello, B.~Zwickl, and S.~Agarwal for useful
discussions. This material is based upon work supported by the Office of Naval
Research under Award Nos. N00014-14-1-0803 and N00014-14-1-0442. ANV thanks the
Institute of Optics for support. LPN acknowledges support from a University of
Rochester Messersmith fellowship.

\onecolumngrid
\newpage
\begin{center}
\textbf{\large Supplemental Materials: Quantum Model of Cooling and Force Sensing With an Optically Trapped Nanoparticle}
\end{center}
\setcounter{equation}{0}
\setcounter{figure}{0}
\setcounter{table}{0}
\setcounter{page}{1}
\makeatletter
\renewcommand{\theequation}{S\arabic{equation}}
\renewcommand{\thefigure}{S\arabic{figure}}
\renewcommand{\bibnumfmt}[1]{[S#1]}
\renewcommand{\citenumfont}[1]{S#1}

\section{Electric fields}
The total electric field as defined in the main article is written as 
\begin{equation}
\label{eq:TotalEField}
\mathbf{E}(\mathbf r,t)=\mathbf{E_{t}}(\mathbf r,t)+\mathbf{E_{p}}(\mathbf r,t)+\mathbf{E_{b}}(\mathbf r,t).
\end{equation}

The $\mathbf{E_t}$ and $\mathbf{E_b}$ fields are both assumed to be Gaussian
beams with the trap treated classically and the probe treated as a quantized
beam with frequency $\omega_{p}$, linewidth $\Delta \omega$, waist $w_{0}$, and
canonical bosonic operators $[a,a^{\dagger}]=1$~\cite{S_Aiello2010}
\begin{equation}
\label{eq:ProbeField}
\mathbf{E_{p}}(\mathbf{r},t)=
i\left(\frac{\hbar \omega_{p}\Delta \omega}{4\pi\epsilon_{0} c}\right)^{1/2}
e^{i\omega_{p}(z/c-t)}\mathbf{e_{p}}G(\mathbf{r'},\omega_{p})a+\text{h.c.},
\end{equation}
where $G(\mathbf r, \omega)$ is the spatial mode function of the beam. The
background field $\mathbf{E_b}$ is simply the quantum field of all other modes
and can be represented via a typical plane-wave expansion~\cite{S_Gerry2004}.

\section{Free field Hamiltonian}
Our configuration Hamiltonian as given in the main text is
\begin{equation}
    H = H_m + H_f + H_\text{int},
\end{equation}
where $H_m = p^2/2m$ is the particle's kinetic energy for the momentum
$p = \sqrt{p_x^2+p_y^2+p_z^2}$, and $H_f$ and $H_\text{int}$ are the free field
and interaction Hamiltonians respectively. The energy of the free field
Hamiltonian is  
\begin{equation}
H_{f}=\epsilon_{0}\int \left|\mathbf{E(r,t)}\right|^{2}\diff3\mathbf{r}. 
\end{equation}
The term proportional to $\abs[0]{\mathbf{E_t(r)}}^2$ corresponds to the energy
of the trap beam, and can be neglected as it represents a constant offset of
the Hamiltonian.
The energy of the free probe field is given by~\cite{S_Aiello2010}
\begin{equation}
    \epsilon_{0}\int\left|\mathbf{E_{p}(r)}\right|^{2}\diff3\mathbf r = \hbar \omega_{p}a^{\dagger}a.
\label{eq:FreeHam}
\end{equation}
The energy of the background field, which can be found in several textbooks
(e.g.~\cite{S_Gerry2004}) can be written in a plane wave expansion as
\begin{equation}
    H_B = \sum_\mu\int \diff3\mathbf k\hbar \omega_{\mathbf k}a^\dagger_\mu(\mathbf k)a_\mu(\mathbf k).
\label{eq:HVacuum}
\end{equation}
The cross term between the trap and background fields, given by
$\int\mathbf{(E_t + E_p)\cdot E_b}\diff3\mathbf r$, vanishes as their mutual
overlap is very small. This cancellation also represents the avoidance of
self-interference and mode overcounting in our model. The cross term between
the probe and the background fields vanishes for the same reason. 
The cross term between the trap and probe fields given by $\mathbf{E_t \cdot
E_p}$, vanishes due to polarization orthogonality. 
Finally, combining Eqs.~(\ref{eq:FreeHam}) and (\ref{eq:HVacuum}), the free field Hamiltonian is
\begin{equation}
\label{eq:FreeFieldHamiltonian2}
H_{f}=\hbar \omega_{p}a^{\dagger}a+\sideset{}{}\sum_{\mu}\int \diff3\mathbf{k}\hbar
\omega_{\mathbf{k}}a^{\dagger}_{\mu}(\mathbf{k})a_{\mu}(\mathbf{k}),
\end{equation}
which is simply the sum of the probe and background field energies.

\section{Interaction Hamiltonian}
The interaction Hamiltonian between the fields and the nanoparticle is given by
\begin{equation}
    H_\text{int}
    = -\frac{1}{2}\int\mathbf{P(r)\cdot E(r)}\diff3\mathbf r.
\end{equation}
Assuming that the particle has a linear polarizability described by
$\mathbf{P(r)} = \alpha_p\mathbf{E(r)}$, the interaction Hamiltonian is
\begin{equation}
    H_\text{int}
    = -\frac{\epsilon_0\epsilon_c}{2}\int_V\abs{\mathbf{E(r)}}^2\diff3\mathbf r,
\end{equation}
where $\epsilon_c = 3(\epsilon_r-1)/(\epsilon_r+2)$
is the Clausius-Mossotti relation for the effective relative permittivity of a
dielectric due to local field effects, and $V$ denotes integration over the
volume of the dielectric particle.

Now when we use the total electric field from Eq.~\eqref{eq:TotalEField} in the
above equation, we again get a number of terms. The terms proportional to
$\abs[0]{\mathbf{E_b}}^2$ and $\abs[0]{\mathbf{E_t \cdot E_p}}^2$ we neglect as
these represent a renormalization of the background modes and we again assume
the trap and probe are cross-polarized.

\subsection{Trap potential}
The first non-negligible term is the effect of the trap beam proportional to
$\int_V\abs[0]{\mathbf{E_p}}^2\diff3\mathbf r$.  Since the nanoparticle is
smaller than the wavelength of any relevant optical field, the integration can
be written as $\int_V\diff3\mathbf r = V\int\delta(\mathbf q)\diff3\mathbf r$,
where $\mathbf q$ is the center of mass position of the particle. Therefore
$H_\text{int}$ for the trap-particle interaction is given by
\begin{equation}
\begin{split}
-\frac{\epsilon_c\epsilon_0}{2}V \abs[0]{\mathbf{E_t(q)}}^2
    &= -\frac{\epsilon_c\epsilon_0}{2}VE_0^2w_0^2\frac{\pi}{2}\abs{G(\mathbf{q})}^2\\
    &= \frac{\epsilon_c\epsilon_0}{2}VE_0^2
        \left(-1
            +\left(\frac{q_z}{z_R}\right)^2+2\left(\frac{q_x}{w_0}\right)^2+2\left(\frac{q_y}{w_0}\right)^2\right)
            +\mathcal{O}(q^4),
\end{split}
\end{equation}
where $q_z$ is the longitudinal coordinate, and $q_x$ and $q_y$ are the
transverse coordinates. Ignoring the overall constant in the above equation,
we finally have for our trap Hamiltonian
\begin{equation}
H_\text{trap} = \frac{p^2}{2m}
    + \frac{1}{2}m \left(\omega_z^2q_z^2 
        + \omega_x^2q_x^2 + \omega_y^2q_y^2
    \right),
\label{eqn:H_trap}
\end{equation}
where $\omega_z = \sqrt{\epsilon_c\epsilon_0E_0^2V/(mz_R^2)}$ and
$\omega_{x,y} = \sqrt{2\epsilon_c\epsilon_0E_0^2V/(mw_0^2)}$. Writing our
canonical nanoparticle variables as
$q_j=\sqrt{\hbar/(2m\omega_j)}(b^\dagger_j+b_j)=\ell_j(b^\dagger_j+b_j)=\ell_jQ_j$ and
$p_j=i\sqrt{\hbar m\omega_j/2}(b^\dagger-b)$ allows us to rewrite
Eq.~\eqref{eqn:H_trap} as $H_\text{trap} = \sum_j\hbar\omega_jb^\dagger_jb_j$.

\subsection{Optomechanical coupling}
Now we examine the coupling between the probe beam and the trapped nanoparticle
given. The Hamiltonian for this term is
\begin{equation}
\begin{split}
H_\text{OM}
    &= -\frac{\epsilon_c\epsilon_0}{2}\int_V \abs[0]{\mathbf{\hat E_p(r)}}^2\diff3\mathbf r\\
    &=-V\frac{\epsilon_c\hbar\omega_p\Delta\omega}{4\pi c}
        \left(a^\dagger a+\frac{1}{2}\right)\abs{G(\mathbf{q-\Delta r})}^2\\
    &=-V\frac{\epsilon_c\hbar\omega_p\Delta\omega}{2\pi^2w_0^2 c}
        \left(a^\dagger a+\frac{1}{2}\right)
        \left(1+2\frac{\Delta z}{z_R^2}q_z +\frac{4}{w_0^2}(\Delta xq_x+\Delta
            yq_y) +\mathcal{O}(\mathbf q^2)\right),
\end{split}
\end{equation}
where we have assumed that the probe beam is shifted from the trap beam by a
small amount $\mathbf{\Delta r} \equiv (\Delta x,\Delta y, \Delta z)$.  Now the
terms proportional to a constant times either $a^\dagger a$ or $q_j$ (for
$j\in\{x,y,z\}$) can be incorporated into shifts in the optical and mechanical
frequencies, and the oscillator position.  The optomechanical coupling term
between the probe and particle is then given by
\begin{equation}
H_\text{OM} = -V\frac{\epsilon_c\hbar\omega_p\Delta\omega}{\pi c}a^\dagger a
\left( \frac{\Delta z}{z_R^2}\ell_zQ_z
    + 2\frac{\Delta x}{z_R^2}\ell_xQ_x
    + 2\frac{\Delta y}{z_R^2}\ell_yQ_y
\right)
= -\sum_j \hbar g_j a^\dagger a Q_j.
\end{equation}

The system Hamiltonian is then a combination of the energies of the probe
field, optical trap, and optomechanical coupling and is given by
\begin{equation}
H_S = \hbar\omega_p a^\dagger a 
    + \sum_j\hbar\omega_jb^\dagger_jb_j
    - \sum_j\hbar g_j a^\dagger a(b_j+b^\dagger_j).
\end{equation}

\subsection{Optical scattering}
Computing the cross coupling Hamiltonian between the field and the background
due to scattering by the nanoparticle gives
\begin{equation}
\begin{split}
-\epsilon_c\epsilon_0\int_V\mathbf{E_t\cdot E_b}\diff3\mathbf r
&\approx -i\frac{\epsilon_cV}{2}\left(\frac{\hbar\epsilon_0}{16\pi^3}\right)^{1/2}
    \int\diff3\mathbf k\sqrt{\omega_k}\sum_\mu
    \left[1 + i\left(\mathbf{k_\perp\cdot q_\perp}+(k_z-k_0)q_z\right)\right]
    \\&\qquad\times
    \left( \mathbf e_\mu (\mathbf k)\cdot \mathbf E_0^*
        a_\mu(\mathbf k) e^{-i(\omega_k-\omega_t)t} + \text{h.c.}
    \right),
\end{split}
\end{equation}
for the trap beam and
\begin{equation}
\begin{split}
-\epsilon_c\epsilon_0\int_V\mathbf{E_p\cdot E_b}\diff3\mathbf r
&\approx -i\frac{\hbar\epsilon_cV}{8\pi^2w_0}
    \sqrt{\frac{2\omega_p\Delta\omega}{\pi c}}
    \int\diff3\mathbf k\sqrt{\omega_k}\sum_\mu
    \left[1 + i\left(\mathbf{k_\perp\cdot q_\perp}+(k_z-k_0)q_z\right)\right]
\\&\qquad\times
    \left(\mathbf{e}_p\cdot \mathbf e_\mu (\mathbf k)a_\mu^\dagger(\mathbf k)a_p
        e^{-i(\omega_p-\omega_k)t)}
        +\text{h.c.}\right)
\end{split}
\end{equation}
for the probe beam, assuming elastic scattering and amplitude of the particle
motion small relative to an optical wavelength~\cite{S_Gieseler2012}. Applying
standard Born-Markov theory and tracing over the background optical modes and
particle motion transverse to the $z$-axis allows us to derive a master
equation
\begin{equation}
\begin{split}
\dot\rho(t) &= -\frac{7\omega_t^5}{\hbar c^6} 
    \frac{(E_0^2\epsilon_0c/2)\epsilon_c^2V^2\ell_z^2}{60\pi}
    \mathcal D [Q_z]\rho(t)
+ \frac{1}{\hbar^4c^3} 
    \frac{(\hbar\omega_p)^4\epsilon_c^2V^2}{24\pi^3w_0^2(c/\Delta\omega)}
    \left(\mathcal D[a_p] +\frac{7\omega_p^2\ell_z^2}{5c^2}\mathcal D[aq_z]\right)\rho(t)
\\&= -\frac{A_t}{2}\mathcal D[Q_z] + \mathcal L_\text{sc}[\rho(t)]\\
\end{split}
\end{equation}

For reference, if we do not trace over the transverse motion, the
three-dimensional analogue of the trap scattering is
\begin{equation}
\mathcal L_t^{3D}\left[\rho(t)\right]
    \equiv -\frac{\omega_t^5}{\hbar c^6} 
        \frac{(E_0^2\epsilon_0c/2)\epsilon_c^2V^2}{60\pi}
        \left( 2\mathcal D[q_\times] + \mathcal D[q_\parallel] + 7\mathcal D[q_z] \right)\rho(t),
\end{equation}
where $\parallel$ and $\times$ indicate transverse motion parallel (or not
parallel) to the polarization direction of the trap beam. A similar expression
holds for the probe
\begin{equation}
\mathcal L_{sc}^{3D}\left[\rho(t)\right]
    \equiv \frac{1}{\hbar^4c^3}
        \frac{(\hbar\omega_p)^4\epsilon_c^2V^2}{24\pi^3w_0^2(c/\Delta\omega)}
        \left\{\mathcal D[a_p] +\frac{\omega_p^2\ell_z^2}{5c^2}
        \left(\mathcal D[aq_\times] + 2\mathcal D[aq_\parallel] + 7\mathcal D[aq_z]\right)\right\}\rho(t).
\end{equation}

\section{Derivation of the detected homodyne current and feedback}
In order to find the input-output relations for the probe field, we write the
system Hamiltonian (for the single degree of freedom $Q_z$) as 
$H_s =
\hbar\omega_p a^\dagger a +\hbar\omega_z b_z^\dagger b_z -\hbar g a^\dagger a
Q_z$. If we move into the interaction picture for the probe field where $a\to
ae^{-i\omega_p t}$, then this becomes $H_s = \hbar\omega_z b_z^\dagger b_z
-\hbar g a^\dagger a Q_z$. 

We assume the probe field is initially a coherent state which can be written as
$a=-i\alpha + v$, where $\alpha$ is a constant and $v$ is a field annihilation
operator. In this case the optomechanical coupling in our system Hamiltonian
becomes
\begin{equation}
H_\text{OM} = -\hbar g_z Q_z a^\dagger a
    = \hbar g_z Q_z \left(\abs{\alpha}^2+i\alpha v^\dagger -i\alpha^* v\right).
\end{equation}
The term proportional to $\abs{\alpha}^2Q_z$ is responsible for simply shifting
the mean position of the oscillator and can safely be ignored, leaving us with
$H_\text{OM} = i\hbar g_z Q_z (\alpha v^\dagger -\alpha^* v)$.

The Heisenberg equation of motion for $v$ is given by
\begin{equation}
\dot v = \frac{1}{i\hbar}[a,H]
    = \alpha g_zQ_z,
\end{equation}
which can be integrated formally to give
\begin{equation}
v(t)= v(t_0) + \int_{t_0}^t\alpha g_zQ_z\dif t' 
    \approx v(t_0) + \alpha g_zQ_z\Delta t,
\end{equation}
where the integration is taken over a time $\Delta t$ short compared to
$1/\omega_z$. If we had picked an initial time $t_f>t$ then we would have
computed $v(t)=v(t_f)-\alpha g_zQ_z\Delta t$. By identifying the input state as
$a_\text{in} = -i\alpha + v(t_0)$, and the output state as
$a_\text{out}=-i\alpha + v(t_f)$, then we can relate the output and input
fields by
\begin{equation}
    a_\text{out} = a_\text{in} + 2\alpha g_z\Delta t
    \equiv a_\text{in} + \frac{\alpha\chi}{2} Q_z,
\end{equation}
where we have defined the variable $\chi\equiv 4g_z\Delta t$.

Now detected homodyne current is proportional to
\begin{equation}
    I_h = \chi^2\Phi\braket{Q_z}(t) + \sqrt{\chi^2\Phi}\xi(t),
\end{equation}
where we have converted from photon numbers to rates by using the identity for
the photon flux $\Phi\equiv\braket{a^\dagger a}\Delta\omega=\alpha^2\Delta\omega$, 
and have introduced the stochastic variable $\xi(t)$ which is due to the shot
noise of the detection.

As described in the main text, phase shifting the measured signal (or
equivalently adding a short time delay) is equivalent to measuring a different
quadrature of motion, i.e. the current that is fed back is $I_\text{fb} =
\chi^2\Phi\braket{P_z}+\sqrt{\chi^2\Phi}\xi'(t)$ , where $\xi'(t)$ has the same
properties as $\xi(t)$. Now writing $\sigma\equiv\chi^2\Phi P_z$ allows us to write
the master equation for the feedback in standard notation~\cite{S_Wiseman1993a}
\begin{equation}
\dot\rho 
    = \mathcal K[\sigma\rho + \rho\sigma^\dagger] + \frac{1}{2\chi^2\Phi}\mathcal K^2[\rho]
    \equiv -i\chi^2\Phi G\left[Q_z^3,\left\{P_z,\rho\right\}\right] -
    \frac{\chi^2\Phi}{2}G^2\mathcal D[Q_z^3]\rho
\end{equation}
where the Liouvillian superoperator $\mathcal K$ is defined as $\mathcal
K[\rho] =[F,\rho]/i\hbar$ and where $F$ is the feedback term that comes from
the feedback Hamiltonian $H_\text{fb} = I_\text{fb} F$, and is chosen to be
$F=\hbar GQ_z^3$ to match the classical physics~\cite{S_Neukirch2015} of the
problem as described in the main text. The gain coefficient $G$ may be related
to the experimental trap beam intensity modulation $M \equiv \Delta I_t/I_t$ by
using the fact that $H_\text{fb} = \hbar\omega_z M Q_z^2 = I_\text{fb}F$, and
therefore~\cite{S_Neukirch2015}
\begin{equation}
    M = \frac{G\chi^2\Phi\braket{Q_z}\!\braket{P_z}}{\omega_z}
    \approx \frac{G\chi^2\Phi N}{\omega_z}.
\end{equation}

\section{Full Master equation}
In the main text of this Letter we presented the master equations describing
the density matrix of only the single degree of freedom represented by the $z$
motion of the nanoparticle. For convenience, we present the full master
equation of the three-dimensional motion of the nanoparticle, as well as of the
probe beam itself, which can be used to calculate the statistics of any quantum
observable for the nanoparticle or probe beam.

The master equation for the density matrix $\rho$ describing the full
three-dimensional motion of the nanoparticle, as well as the probe field is
given by
\begin{equation}
\begin{split}
\dot\rho(t)
&= \frac{1}{i\hbar}\left[H_S,\rho(t)\right]
        \quad\bigg\}\text{ Unitary dynamics}
\\ &\qquad 
    + \mathcal L_t^{3D}\left[\rho(t)\right]
    + \mathcal L_\text{sc}^{3D}\left[\rho(t)\right]
        \quad\bigg\}\text{ Photon scattering}
\\ &\qquad 
    -\frac{\eta_fk_BT}{\hbar^2}\left[\mathbf q,\left[\mathbf q,\rho(t)\right]\right]
    -\frac{\eta_f}{12k_BTm^2}\left[\mathbf p,\left[\mathbf p,\rho(t)\right]\right]
    -\frac{\eta_f}{2m\hbar}\left[\mathbf q,\left\{\mathbf p,\rho(t)\right\}\right]
        \quad\bigg\}\text{ Gas scattering}.
\end{split}
\end{equation}
This equation can be used for studying full dynamics, light-matter entanglement
and photon statistics.

\section{Homodyne photodetection: a comparison of theory and experiment}

The experimental data in Fig.~2 of the main article was obtained by using a
scheme more involved, and of more general applicability, than indicated in the
simplified detection model presented in the main article. In the laboratory,
the position of the nanoparticle was determined from the optical interference
between the unscattered probe and a spherical wave radiated by the induced
dipole of the polarizable sphere. The interference signal varies linearly with
the particle position for oscillation amplitudes small compared to the optical
wavelength. In order to detect the particle position and also to eliminate the
large constant background term (equal to the unscattered probe flux
$\approx\Phi$) in an experimental setting, we leveraged a balanced homodyne
detection scheme, as described in~\cite{S_Gieseler2012}
and~\cite{S_Neukirch2015}.

In this configuration, one of the detectors sampled the entire spatial profile
of the probe beam, while the other was apodized so that only the center of the
beam was sampled. As the trapped particle moved in the axial direction, the
relative amount of scattered light collected by each detector, and thus the AC
term of the resulting homodyne current, was modulated. The optical channels
were adjusted so that each detector recorded the same average power, and the
resulting homodyne signals were subtracted to eliminate the common DC bias.

In our theoretical model, rather than include an additional quantized mode
corresponding to the full induced dipole emission pattern, we considered for
simplicity only the field scattered back into the probe mode by the
nanoparticle. This simplification is admissible as long as the value of the
linear optomechanical coupling constant $\chi$ in the model is taken from
experiment~\cite{S_Gieseler2012}. We emphasize that our experimental method of
feedback cooling works even if the trap and probe foci are very close to each
other, in which case the optomechanical coupling is essentially quadratic in
the particle position, while the position detection (using the dipole wave
mode) is still linear.  In the theoretical calculation of the phonon number,
the linear (or quadratic) optomechanical coupling simply adds a position offset
(or frequency shift) to the trap. Both the offset as well as the shift are
negligible for weak probe light used in the experiment, rendering the
theoretical predictions of ground state cooling identical for linear as well as
quadratic optomechanical coupling.

\section{Computing the gas damping rate as a function of gas pressure}
The damping rate of the particle motion due to the background gas is given by
$\Gamma_0 \equiv \eta_f/m$, where $m$ is the particle mass and $\eta_f =
6\pi\mu r_d$ is the coefficient of friction for a spherical particle of radius
$r_d$ in a fluid of dynamic viscosity $\mu$. This expression is for a particle
much larger than the mean free path of the gas bath $\lambda_\text{mfp}$.
For a nanoparticle in a rarefied gas, as assumed in Eq.~(9) in the main text,
the damping constant becomes~\cite{S_Beresnev2006} 
\begin{equation}
    \Gamma_0 = \frac{6\pi\mu r_d}{m}\times \text{correction}
    = \frac{6\pi\mu r_d}{m} \frac{0.619}{0.619 + \text{Kn}}\left(1+\frac{0.31\text{Kn}}{0.785+1.152\text{Kn}+\text{Kn}^2}\right),
    \label{eqn:Gamma0}
\end{equation}
where the correction term is in terms of the Knudsen number $\text{Kn}\equiv
\lambda_\text{mfp}/r_d \propto 1/Pr_d$.  Using the fact that
$\lambda_\text{mfp}\approx\SI{70}{nm}$ at atmospheric pressure, and matching
the experimentally measured rates~\cite{S_Gieseler2012,S_Neukirch2015}, we can
write the damping for a general pressure and particle size as
\begin{equation}
\Gamma_0 \approx \frac{r_d}{\SI{70}{nm}}\frac{2\pi\times\SI{e6}{Hz}}{0.619 + \text{Kn}}\left(1+\frac{0.31\text{Kn}}{0.785+1.152\text{Kn}+\text{Kn}^2}\right).
\end{equation}

In order to ensure that this model of damping holds in the deep quantum
regime~\cite{S_Hornberger2004}, we consider the effects of individual
scattering events as described by the master equation~\cite{S_Diosi1995}
\begin{equation}
    \dot\rho = \sum_V V\rho V^\dagger - \frac{1}{2}\left\{V^\dagger V,\rho\right\},
\end{equation}
where the Lindblad operators are given by
\begin{equation}
    V\sim \exp\left(-i\frac{\Delta p q}{\hbar}\right),
\end{equation}
where $\Delta p$ is the momentum transfered to the nanoparticle during a
collision with a gas molecule. Now if the momentum kick is sufficiently small,
then we can approximate
\begin{equation}
    \exp\left(-i\Delta pq/\hbar\right) \approx 1 - i\Delta pq/\hbar, 
    \label{eqn:gaskick}
\end{equation}
which ultimately allows us to recover the Brownian contribution, derived
in~\cite{S_Diosi1995}, to our master equation presented in the main Letter. In
order for the approximation of Eq.~\eqref{eqn:gaskick} to hold, it is necessary
that $\Delta p \ell_\text{coh}/\hbar < 1,$ where $\ell_\text{coh}$ is the
coherence length of the system. Considering the coherence length of the system
in its ground state, $\ell_\text{coh} \approx \ell_z$. Using the parameters
listed in the main article for the system at $T = \SI{4}{K}$, and using the
maximum RMS thermal momentum change $\Delta p = 2\sqrt{3m_\text{gas}k_B T}$
assuming a gas made of $\text{N}_2$ gives
\begin{equation}
    \frac{\Delta p\ell_z}{\hbar} \approx 0.6 < 1,
\end{equation}
which ensures that our linearized treatment of Brownian motion is valid over
all considered regimes. This inequality actually overestimates the magnitude of
a typical momentum kick as Hydrogen and Helium comprise the majority of the
species of gas molecules in an environment at \SI{4}{K}, and are much lighter
than $\text N_2$.

\section{Converting a master equation to a Langevin equation}
Consider a master equation in standard Linblad form
\begin{equation}
    \dot\rho = \mathcal L_0\rho = \frac{1}{i\hbar}\left[H,\rho\right] - \sum_j\frac{\gamma_j}{2}\mathcal D[L_j]\rho,
\end{equation}
where $H_0$ is a Hamiltonian representing the unitary evolution and $\mathcal
D[L_j]\rho$ is the standard Lindblad dissipation superoperator for the Lindblad
operators $L_j$~\cite{S_Lindblad1976}.  Now following the quantum-state
diffusion model presented by Gisin and Percival~\cite{S_Gisin1992}, the
density matrix can be considered as the expectation over pure states e.g. $\rho
\equiv \E{\ketbra{\psi}{\psi}}$, whose dynamics are stochastic in nature. The
(stochastic) differential increment for the equation of motion for $\ket{\psi}$
representing the unraveling of the master equation given by the QSD model
is~\cite{S_Gisin1992, S_Halliwell1995}
\begin{equation}
\ket{\dif\psi}
= \frac{1}{i\hbar}H\dif t\ket{\psi}
    + \sum_j\bigg[
          \frac{\gamma_j}{2}\left(2\braket{L_j^\dagger}L_j - L_j^\dagger L_j - \braket{L_j^\dagger}\braket{L_j}\right)\dif t
        + \left(L_j - \braket{L_j}\right)\sqrt{\gamma_j}\dif W_j
    \bigg]\ket{\psi}
\label{eqn:SSE}
\end{equation}
where $\dif W_j$ are complex differentials representing Wiener processes
defined by $\E{\dif W_j} = \E{\dif W_j^*} = \E{\dif W_j\dif W_k} = 0$, and
$\dif W_j^*\dif W_k = \delta_{jk}\dif t$. Note, this complex Wiener process is
equivalent to $\dif W = (\dif R_1 + i\dif R_2)/\sqrt{2}$, where $\dif R_j$ are
real valued Wiener increments. Now such a model is not unique, but represents
an open quantum system which is continuously monitored by it's
environment~\cite{S_Brun1997}, and thus provides an appropriate trajectory or
unraveling for a system that we will ultimately wish to observe in terms of a
PSD or measurement spectrum (i.e.~which we will continuously monitor).

Now we are interested in expressing the random forces acting on our system due
to interaction with the environment (as expressed by the Lindblad operators
$L_j$). Therefore computing the increment for the momentum $p$ of the system
gives
\begin{equation}
\begin{split}
\dif p 
    &= \braket{\dif\psi|p|\psi} + \braket{\psi|p|\dif\psi} + \braket{\dif\psi|p|\dif\psi} + \mathcal O(\dif t^2)\\
    &= -\frac{1}{i\hbar}\left[H_0,p\right]\dif t -\sum_j\frac{\gamma_j}{2}\mathcal D\left[L_j^\dagger\right]p\dif t + \sum_j\left(\dif B^\dagger_j p + p\dif B_j\right)\\
    &= \mathcal L_0^\#[p]\dif t + \sum_j\left(\dif B^\dagger_j p + p\dif B_j\right),
\end{split}
\end{equation}
where $\dif B_j \equiv (L_j - \braket{L_j})\sqrt{\gamma_j}\dif W_j$. Applying
this to our master equation gives us our Langevin equation for the stochastic
forces listed in Eq.~(15) in the main article.

\end{document}